# Complex Response Function
# of Magnetic Resonance Spectrometers

G. Annino[@], M. Cassettari, M. Fittipaldi[#], and M. Martinelli

*Istituto di Fisica Atomica e Molecolare\* del CNR, via G. Moruzzi 1, 56124 Pisa (Italy)*
[#]*now with Huygens Laboratory, University of Leiden, P.O. 9405, 2300 RA Leiden (The Netherlands)*

\*now *Istituto per i Processi Chimico-Fisici*
[@]e-mail: geannino@ifam.pi.cnr.it

(August 27, 2002)

## Abstract

A vectorial analysis of magnetic resonance spectrometers, based on traveling wave resonators and including the reference arm and the automatic control of frequency, has been developed. The proposed model, valid also for stationary wave resonators, gives the response function of the spectrometer for any working condition, including scalar detectors with arbitrary response law and arbitrary excitation frequency. The purely dispersive and purely absorptive linear responses are discussed in detail for different scalar detectors. The developed approach allows for optimizing the performances of the spectrometer and for obtaining the intrinsic lineshape of the sample in a very broad range of working conditions. More complex setups can be modeled following the proposed scheme.

## I. Introduction

Resonant structures based on traveling wave modes have been widely used in different fields of physics; a more and more popular example is represented by dielectric resonators working on whispering gallery modes [1, 2, 3]. Among the different applications, an innovative realization of magnetic resonance spectrometers based on dielectric resonators has been recently proposed, specifically useful for high magnetic field applications [4]. An effective employment of such devices requires otherwise an accurate modeling of the resonator and of the overall apparatus.

The basic aim of this paper, which generalizes a previous study [5], is the theoretical investigation under general hypotheses of the response of this class of spectrometers, including the reference arm and the automatic control of frequency. In particular the complex response of the spectrometer will be calculated by expressing amplitude and phase of the radiation in terms of its basic parameters. This approach appears conceptually more appropriate than the usual RLC lumped circuit representation, since it is based on quantities measurable at any frequency (including the quasi optical limit). The complex formulation is furthermore justified by the availability of vectorial analyzers working up to the THz region [6].

The developed analysis will focus on the relevant aspects of the coherent response of the spectrometer (i.e. the response calculated neglecting any noise effects), in which the resulting signal is completely defined by the parameters controlled by the operator. The effects of the different sources of noise, which are beyond the scope of this work, can be included *a posteriori*.

The generality of the proposed approach, based on the complex resonance frequency, the free spectral range and the coupled energy, makes inessential any further details on the resonator. The obtained results can thus be applied to spectrometers based on other kinds of resonators, such as metallic cavities and Fabry-Perot resonators; indeed, some important findings obtained in the works of Feher [7] and Wilmshurst et al. [8] will be reproduced in this paper.

The effect of a sample interacting with the resonator will be developed, for sake of simplicity, by using a perturbative approach. To this purpose a generalization of the celebrated Boltzmann-Ehrenfest theorem [9] is used. The above procedure always leads to a linear behavior, which can be considered as the intrinsic response of the spectrometer. On the other hand, the use of scalar detectors can lead to a non-linear response in specific conditions.

The application of the obtained results to a typical working condition will be discussed and experimentally verified in a forthcoming paper.

The plan of the paper is the following. Section II discusses the main properties of a traveling wave resonator coupled to a transmission line, in presence also of a reference arm. In Section III the effects of a perturbation of the resonator in terms of the variation of its complex resonance frequency are obtained. In Section IIIa the perturbed resonance frequency is derived in terms of magnetic permeability by means of a generalization of the Boltzmann-Ehrenfest theorem. The vectorial response of the magnetic spectrometer is discussed in Section IV. The obtained results are specialized to the case of scalar detectors with arbitrary response law in Section V. In Section VI the absorptive response in linear regime is discussed in detail for different scalar detectors. Finally, concluding remarks are given in Section VII.

## II. General background

The vectorial modelization of a traveling wave resonator, coupled to a general transmission line in the reaction configuration, can be effectively developed in terms of a circuit formed by a directional coupler and a phase shifter [10], as schematically shown in Fig. 1.

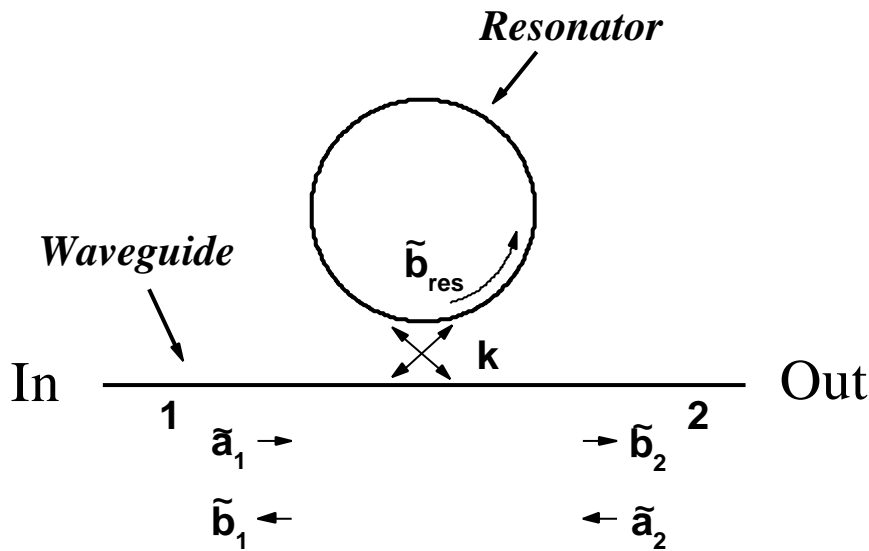

**Fig. 1.** *Scheme of a traveling wave resonator coupled to a transmission line.*

Here $\tilde{a}_j$ and $\tilde{b}_j$ represent the incident and reflected complex wave amplitudes at the reference plane **j**, normalized in order that $|\tilde{a}_j|^2$ and $|\tilde{b}_j|^2$ give the powers flowing inwards and outwards the relative reference planes, respectively. The relations among the amplitudes at the different reference planes can be obtained by using the method of the scattering matrix [10, 11]. In order to have a purely traveling wave in the resonator, port 2 is terminated with a matched load, so that $\tilde{a}_2 = \tilde{b}_1 = 0$; the forward wave and the wave coupled to the resonator can now be conveniently expressed as

$$b = \frac{\tilde{b}_2}{\tilde{a}_1} = \frac{\sqrt{1-k^2} - e^{-J}}{1 - \sqrt{1-k^2} \times e^{-J}} \qquad (1)$$

and

$$b_{res} = \frac{\tilde{b}_{res}}{\tilde{a}_1} = \frac{i \times k}{1 - \sqrt{1-k^2}\, e^{-J}} \qquad (2),$$

respectively. In these expressions the complex wave attenuation $J$ is given by $J = a + i \times j$, where $a$ is the attenuation of the wave and $j$ its phase shift in a round trip, while $k$ is the coupling coefficient of the transmission line to the resonator [10]; $i$ represents the imaginary unit. According to this model, the critical coupling coefficient $k_c$ (given by the condition $b = 0$) is obtained when

$$\sqrt{1 - k_c^2} = e^{-a} \qquad (3),$$

where $j = 2\pi n$ has been taken into account for resonances of order $n$.

If the finesse $\mathcal{F}$ of the resonator satisfies the condition $\mathcal{F} >> 1$, which implies $a << 1$, the loaded merit factor $Q_L$ can be explicitly calculated [10, 12]. In the limit of weak coupling condition, i.e. $k^2$ of the order of $a$, the loaded merit factor results

$$Q_L = \frac{n\pi}{a + \frac{k^2}{2}} \qquad (4).$$

In particular all the coupling conditions up to the critical one (and beyond) can be described in this approximation. The following analysis will be developed according to the described framework, which includes indeed all the usual working conditions [12]. The effects of the stationary waves that can be induced by the sample will be neglected.

It is worthwhile to note that the proposed model also applies to conventional standing wave resonators excited in reflection configuration, when the signal $\tilde{a}_2$ entering the port 2 is disregarded. Indeed, in this case the wave amplitudes $\tilde{a}_1$ and $\tilde{b}_2$ represent the incoming and the reflected wave respectively, and the scattering matrix preserves the same form which results only from the energy conservation principle (Chap. 4 of Ref. 13). Accordingly, the proposed model will be valid for a wide class of resonant structures. Moreover, the model can be generalized in order to include the reference arm; the related circuit is sketched in Fig. 2.

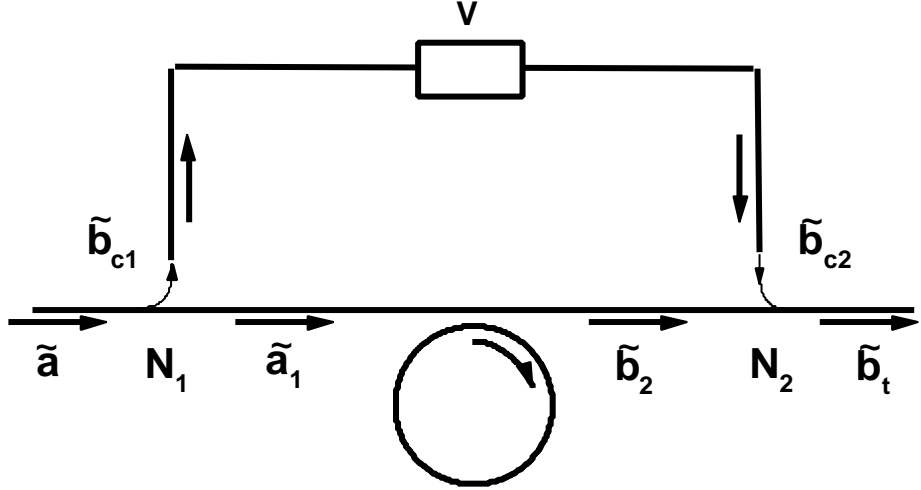

**Fig. 2.** *Scheme of a spectrometer based on a traveling wave resonator and including a reference arm; the nodes $N_1$ and $N_2$ act as directional couplers. The device labeled with **V** allows to change amplitude and phase of the incoming wave.*

The forward amplitude $\tilde{b}_2$ can now be expressed by $\dfrac{\tilde{b}_2}{\tilde{a}} = \dfrac{\tilde{b}_2}{\tilde{a}_1} \times \sqrt{1 - \dfrac{|\tilde{b}_{c1}|^2}{|\tilde{a}|^2}}$. The total forward amplitude $\tilde{b}_t$ can then be expressed as

$$\frac{\tilde{b}_t}{\tilde{a}} = \frac{\tilde{b}_2}{\tilde{a}_1} \times \sqrt{1 - \frac{|\tilde{b}_{c1}|^2}{|\tilde{a}|^2}} + \frac{\tilde{b}_{c2}}{\tilde{a}} \qquad (5),$$

where $\tilde{b}_{c2}$ enters into the node $N_2$ from the reference arm. By using the definitions $b_t = \dfrac{\tilde{b}_t}{\tilde{a}}$, $b_{c1} = \dfrac{\tilde{b}_{c1}}{\tilde{a}}$, $b_{c2} = \dfrac{\tilde{b}_{c2}}{\tilde{a}}$, Eq. 5 becomes

$$b_t = b \sqrt{1 - |b_{c1}|^2} + b_{c2} \qquad (5a).$$

Eq. 5a represents a generalization of Eq. 1 for a typical experimental configuration.

### III. Perturbed resonator

The theoretical analysis of the response of a magnetic resonance spectrometer employing a vectorial detector can be reduced to the analysis of the circuits described in Fig. 1 and in Fig. 2. The presence of a sample modifies the parameters of the resonator when the magnetic resonance is swept across; in turn, these modifications change the amplitude of the forward wave, as described by Eq. 1 or Eq. 5a. The knowledge of the complex response function of the spectrometer is then traced back to the knowledge of the effects of the magnetic resonance on the parameters of the resonator. To this purpose the analysis will start by the description of the basic circuit represented in Fig. 1; the effects of the reference arm will be considered later.

In the most general approach, the knowledge of the complex magnetic permeability $\hat{m}(B) = m' - i \times m''$ of the sample allows the determination of the complex resonance frequency $\hat{\omega} = \omega' + i \times \omega''$ of the resonator (according to the procedure reported in Ref. 14). From $\hat{\omega}$ and the related fields distribution the complex attenuation $J(B)$ and the coupling coefficient $k(B)$ can be calculated. Assuming that the electromagnetic field distribution is not affected by the magnetic resonance, only the first term of the perturbative expansion of the complex response can be retained; the obtained behavior can then be considered as the intrinsic response of the spectrometer. The variation of $b$ around the starting point $b_0$ thus becomes

$$b(B) = b_0 + \Delta b(B)$$

with

$$\Delta b(B) = \frac{\partial b}{\partial J} \Delta J(B) \qquad (6),$$

having assumed negligible variations of the coupling intensity $k^2$ across the magnetic resonance.

Since the derivative

$$\frac{\partial b}{\partial J} = \frac{k^2 e^{-J}}{\left(1 - \sqrt{1 - k^2 e^{-J}}\right)^2} = -b_{res}^2 \times e^{-J} \qquad (7)$$

is nonvanishing for any finite $k$ and $\alpha$, the first order analysis of the complex response of the system is valid for any allowed starting point. As a consequence, the parametric curves $\Delta b(B)$ and $\Delta J(B)$ have the same shape in the complex field.

For the resonance with modal index $n$ and frequency $\omega_n'$, the variation $\Delta J$ can be determined from the relation $J = \alpha + i \times \left(2\pi \frac{\omega - \omega_n'}{\omega_{fsr}} + 2\pi n\right) = \alpha + i \times (\delta\varphi + 2\pi n)$; here $\delta\varphi$ is the phase mismatch due to the difference between $\omega_n'$ and the excitation frequency $\omega$, and $\omega_{fsr}$ is the free spectral range of the resonator around $\omega_n'$. In particular, the condition $\hat{Q} >> 1$ leads to $\delta\varphi << 1$. Assuming $\omega_n' = n \times \omega_{fsr}$ it follows

$$\Delta J = \Delta\alpha + i \times 2\pi n \left(-\frac{\omega}{\omega_n'^2} \Delta\omega_n'\right) = \Delta\alpha - i \times 2\pi n \frac{\Delta\omega_n'}{\omega_n'} \qquad (8)$$

where the term $\frac{\omega}{\omega_n'^2} \Delta\omega_n'$ has been approximated to $\frac{\Delta\omega_n'}{\omega_n'}$ [15].

On the other hand, the complex resonance frequency can be expressed as $\hat{\omega}_n = \omega_n' + i \times \frac{\omega_n'}{2 \times Q_L}$ (Chap. 5 of Ref. 13); its variation at the first order is

$$\Delta\hat{\omega}_n = \Delta\omega_n' + i \times \omega_n' \frac{1}{2\pi n} \Delta\alpha \qquad (9).$$

Eqs. 8 and 9 give $\Delta J = \left(-i \times \frac{2\pi n}{\omega_n'}\right) \Delta\hat{\omega}_n$, and thus Eq. 6 can be rewritten as

$$\Delta b(B) = -i \times \frac{2\pi n}{\omega_n'} \frac{\partial b}{\partial J} \Delta\hat{\omega}_n(B) \qquad (10).$$

The response function of the system is then related to the variation of the complex resonance frequency $\Delta\hat{\omega}_n(B)$. To be consistent with the previous analysis, also $\Delta\hat{\omega}_n(B)$ will be calculated on the basis of a perturbative approach.

A variable excitation frequency can be included in the proposed model coming back to the variation of the complex attenuation $J$, now given by

$$\Delta\vartheta = \Delta\alpha - i \times 2\mu \left( \frac{\Delta\omega'_n - \Delta\omega'}{\omega'_n} \right).$$

If the excitation frequency follows the resonance frequency, as occurs when an Automatic Frequency Control (AFC) system is used, the imaginary part of $\Delta\vartheta$ is cancelled out and the complex attenuation equals $\Delta\alpha$.

### IIIa Generalized Boltzmann-Ehrenfest theorem

The frequency variation induced by a perturbation in a system oscillating at the frequency $\omega'$ can be related to the variation of its mean energy $\langle W \rangle$ by using the Boltzmann-Ehrenfest theorem, which ensures that [9]

$$\frac{\Delta \langle W \rangle}{\langle W \rangle} = \frac{\Delta \omega'}{\omega'}.$$

If an ideal lossless electromagnetic system (ideal resonator) is considered, the change of its mean stored energy is due to the variation of the (real) magnetic permeability $\Delta\mu$. When this variation is related to a gyrotropic susceptibility $\ddot{\chi} = \frac{\Delta \ddot{\mu}}{4\pi}$, as usual for magnetic systems in a static magnetic field $\mathbf{B}$, $\ddot{\chi}$ is given by (Chap. 8 of Ref. 13)

$$\ddot{\chi} = \begin{pmatrix} \kappa & -i\times\nu & 0 \\ i\times\nu & \kappa & 0 \\ 0 & 0 & 0 \end{pmatrix}$$

with $\kappa, \nu \in \mathbb{R}$.

The variation of the mean energy stored in the resonator, calculated at the first order, can be expressed as [16]

$$\Delta \langle W \rangle = -\frac{1}{2} \int_{sample} \left[ (\kappa + \nu)\langle H_+^2 \rangle + (\kappa - \nu)\langle H_-^2 \rangle \right] dV,$$

where $H_+$ and $H_-$ are the components of the magnetic field of the normal mode, rotating and counter-rotating in comparison to the precession induced by the static magnetic field, respectively.

Neglecting as usual the contribution of the counter-rotating component, the fractional change of energy can be expressed as

$$\frac{\Delta \langle W \rangle}{\langle W \rangle} = -\frac{1}{2} \frac{\Delta \mu'}{\mu'} \frac{\int_{sample} \mu' \langle H_+^2 \rangle dV}{\int_{resonator} \mu'(r) \langle H^2 \rangle dV} = -\frac{1}{2} \frac{\Delta \mu'}{\mu'} \eta,$$

where $\mathbf{H}$ is the total magnetic field of the resonant mode, $\Delta\mu'$ the variation of the rotating component of the permeability tensor, and $\eta$ the related magnetic filling factor; the initial magnetic permeability $\mu'$ is considered as a scalar homogeneous quantity. From the Boltzmann-Ehrenfest theorem it follows then that $\frac{\Delta\omega'}{\omega'} = -\frac{1}{2}\eta \frac{\Delta\mu'}{\mu'}$.

When the magnetic losses of the sample are taken into account, the merit factor of the resonator becomes $Q^{-1} = 2\frac{\omega''}{\omega'} = \eta \frac{\mu''}{\mu'}$ so that, at the first order, $\frac{\Delta\omega''}{\omega'} = \frac{1}{2}\eta \frac{\Delta\mu''}{\mu'}$. The Boltzmann-Ehrenfest theorem can then be generalized to lossy systems as follows:

$$\frac{\Delta\hat{\omega}}{\omega'} = -\frac{1}{2}\eta \frac{\Delta\hat{\mu}}{\mu'} \qquad (11).$$

When different sources of losses are present the above derivation can be generalized, provided that the fields distribution is still unchanged; indeed in this case the losses can be superposed linearly.

In conclusion, from Eqs. 10 and 11 it results

$$\Delta b(B) = i \times 4\pi^2 n \frac{h}{mc^2} \frac{\partial b}{\partial J} \hat{\chi}(B) \quad (12),$$

where the complex susceptibility for the rotating component $\hat{\chi} = \chi' - i \times \chi''$ has been introduced.

Under the previous assumptions, the shape of $\Delta b(B)$ induced by the magnetic resonance accurately reproduces the intrinsic complex lineshape of the magnetic susceptibility $\hat{\chi}(B)$; on the contrary, the resulting lineshape obtained when scalar detectors are employed can be completely distorted, as shown in Sect. 5.

### IV. Vectorial Response

The above analysis shows that the complex response function of the system represented in Fig. 1 is encoded in the complex quantity

$$b(B) = b_0 + i \times 4\pi^2 n \frac{h}{mc^2} \frac{\partial b}{\partial J} \hat{\chi}(B) \quad (13).$$

When the excitation frequency $\omega$ is varied around $\omega_n'$, in absence of magnetic resonance the amplitude $b_0$ describes the resonance curve of the mode of index $n$, represented in the complex plane $(\text{Re}[b], \text{Im}[b])$ by a circumference [12]. The choice of the excitation frequency fixes the phase shift $\varphi$ and then the starting point $b_0$.

The more general configuration represented in Fig. 2 can be analyzed combining Eq. 5a with Eq. 13, which give

$$b_t(B) = b_0 \times \sqrt{1 - |b_{c1}|^2} + b_{c2} + i \times 4\pi^2 n \frac{h}{mc^2} \frac{\partial b}{\partial J} \sqrt{1 - |b_{c1}|^2} \hat{\chi} = b_0^t + \Delta b_t \quad (14),$$

where $b_0^t = b_0 \times \sqrt{1 - |b_{c1}|^2} + b_{c2}$ is the effective starting point, and $\Delta b_t = \Delta b \times \sqrt{1 - |b_{c1}|^2}$.

The complex response of the spectrometer is given in general by a linear superposition of the real and imaginary components of $\hat{\chi}$, unless $\frac{\partial b}{\partial J}$ has only one nonvanishing component. The analysis of $\frac{\partial b}{\partial J}$ is detailed in App. 1.

The ratio between the response $\Delta b_t(B)$ and $\hat{\chi}(B)$ can be defined as the complex amplification $A(J,k)$ of the spectrometer, which results (see also Eq. 7)

$$A(J,k) = i \times 4\pi^2 n \frac{h}{mc^2} \frac{\partial b}{\partial J} \sqrt{1 - |b_{c1}|^2} = -i \times 4\pi^2 n \frac{h}{mc^2} e^{-J} (b_{res})^2 \sqrt{1 - |b_{c1}|^2} \quad (15).$$

Taking into account that $|b_{res}|^2 = \frac{1 - |b|^2}{1 - e^{-2\alpha}}$, as follows from Eqs. 1 and 2, the modulus of $A(J,k)$ results

$$|A(J,k)| = 4\pi^2 n \frac{h}{mc^2} \frac{e^{-\alpha}}{1 - e^{-2\alpha}} \left(1 - |b_0|^2\right) \sqrt{1 - |b_{c1}|^2} \quad (16).$$

At the first order in $\alpha$, $|A(J,k)|$ becomes

$$|A(J,k)| = \frac{2\pi}{\mu c} \sqrt{1 - |b_{c1}|^2} \left(1 - |b_0|^2\right) Q_0 \eta = \frac{2\pi}{\mu c} \sqrt{1 - |b_{c1}|^2} P_d Q_0 \eta \qquad (17),$$

where $P_d = 1 - |b_0|^2$ is the fraction of the power $|\tilde{a}_1|^2$ dissipated in the resonator. $|A(J,k)|$ is then proportional to the *unloaded* merit factor $Q_0 = \frac{\nu \pi}{\alpha}$, and reaches its maximum value at the critical coupling condition. The complex amplification $A$, related to the coherent response of the spectrometer, is then reduced by the presence of a reference arm owing to the reduced power available to the resonator.

As the magnetic resonance experiments are in general performed by using scalar detectors, the scalar response of the spectrometer will be analyzed in detail in the next section. In this case, in spite of the linearity of the complex signal $b_t(B)$ in terms of the magnetic susceptibility, a non linear response can arise owing to the non linearity of the employed detector.

### V. Scalar response

In order to allow a compact analysis of the scalar response of the spectrometer it is useful to introduce a vector formalism in which a complex number $c$ corresponds to a 2-dimensional vector $\vec{c} = \begin{pmatrix} \mathrm{Re}[c] \\ \mathrm{Im}[c] \end{pmatrix}$; in this formalism $\vec{c}^{\wedge} \stackrel{\text{def}}{=} \begin{pmatrix} -\mathrm{Im}[c] \\ \mathrm{Re}[c] \end{pmatrix}$. Eq. 14 can now be rewritten in terms of vectors as

$$\vec{b}_t(B) = \vec{b}_0^t + \begin{pmatrix} \vec{A} \times \vec{c} \\ \vec{A} \times \vec{c}^{\wedge} \end{pmatrix} = \vec{b}_0^t + \begin{pmatrix} c' & c'' \\ -c'' & c' \end{pmatrix} \begin{pmatrix} \mathrm{Re}[A] \\ \mathrm{Im}[A] \end{pmatrix}$$

where $\vec{A} \times \vec{c}$ is the scalar product between the amplification vector $\vec{A}$ and the vector $\vec{c} \stackrel{\text{def}}{=} \begin{pmatrix} c' \\ c'' \end{pmatrix}$.

The scalar response of the spectrometer can be investigated by introducing the generalized amplitude of the wave $\vec{b}_t$, defined as

$$\left|\vec{b}_t(B)\right|^r = \sqrt[r]{\left|\vec{b}_0^t\right|^2 + \left|\begin{matrix} \vec{A} \times \vec{c} \\ \vec{A} \times \vec{c}^{\wedge} \end{matrix}\right|^2 + 2\vec{b}_0^t \times \begin{pmatrix} \vec{A} \times \vec{c} \\ \vec{A} \times \vec{c}^{\wedge} \end{pmatrix}} \qquad (18),$$

where the exponent $r > 0$ takes into account the responsivity of the detector.

In general the analysis of the scalar response of the system requires a direct use of the basic Eq. 18.

From a conceptual point of view the condition of ideal compensation $\vec{b}_0^t = 0$ appears relevant; in this case the overall forward wave is just given by the contribution of the magnetic resonance, and Eq. 18 reduces to

$$\left|\vec{b}_t(B)\right|^r = \left(|\vec{A}| \times |\vec{c}|\right)^r \propto \left(c'^2 + c''^2\right)^{\frac{r}{2}};$$

the obtained lineshape thus non-linearly mixes the dispersive and absorptive components of $\hat{c}$. This mixing can be avoided by using the AFC system, which cancels out the contribution of $c'$; the resulting signal, proportional to $|c''|^r$, is still distorted in comparison to $c''$ excluding the case $r = 1$. However, when the response curve crosses the critical coupling point distorted lineshapes are in general obtained also for linear detector, as pointed out also by Feher [8].

Further interesting cases occur when the condition $|\Delta \vec{b}_t(B)| \ll |\vec{b}_0^t|$ is satisfied, that is when the magnetic resonance contribution to the forward wave represents a weak correction to the overall signal. In this case only the linear terms in $\chi'$ and $\chi''$ can be maintained in the expansion of $|\vec{b}_t|^r$, so that the previous condition represents the linear regime of the spectrometer. The generalized amplitude reduces now to

$$|\vec{b}_t(B)|^r \approx \sqrt[r]{|\vec{b}_0^t|^2 + 2\vec{A} \times \vec{b}_0^t \cdot \chi' + 2\vec{A} \times \vec{b}_0^{t\,\wedge} \cdot \chi''} \approx |\vec{b}_0^t|^r + r|\vec{b}_0^t|^{r-2} \vec{A} \times (\vec{b}_0^t \cdot \chi' + \vec{b}_0^{t\,\wedge} \cdot \chi'') \quad (19).$$

The signal $\Delta|\vec{b}_t|^r = |\vec{b}_t|^r - |\vec{b}_0^t|^r$ then arises from a linear mixing between $\chi'$ and $\chi''$, defined by the projections of the amplification $\vec{A}$ on the starting vectors $\vec{b}_0^t$ and $\vec{b}_0^{t\,\wedge}$. This mixing can be determined by defining a 'mixing angle' $\xi$ between the vectors $\vec{b}_0^t$ and $\vec{A}$ (that is between $\vec{b}_0^t$ and $\frac{\partial \vec{b}}{\partial J}^{\wedge}$); accordingly, $\Delta|\vec{b}_t|^r$ becomes

$$\Delta|\vec{b}_t(B)|^r = r |\vec{b}_0^t|^{r-1} \times |\vec{A}| (\cos\xi \times \chi' + \sin\xi \times \chi'') \quad (20).$$

Eq. 20 shows when this inherent mixing can be avoided (besides the already discussed case of external AFC). In particular when $\xi = 0$ or $\xi = \pi$, that is $\frac{\partial \vec{b}}{\partial J} \wedge \vec{b}_0^t$, the linear contribution of $\chi''$ disappears and the response of the system is purely dispersive. In this case the signal becomes

$$\Delta|\vec{b}_t(B)|^r = r \times \text{sgn}[\vec{A} \times \vec{b}_0^t] \times |\vec{b}_0^t|^{r-1} |\vec{A}| \chi' \quad (21).$$

The sensitivity of the spectrometer can now be characterized by introducing the scalar amplification $A_r'(J,k) = r \times \text{sgn}[\vec{A} \times \vec{b}_0^t] \times |\vec{b}_0^t|^{r-1} |\vec{A}|$.

If the reference arm is not used, the condition $\frac{\partial \vec{b}}{\partial J} \wedge \vec{b}_0^t$ is fulfilled when the equation

$$\cos(\delta\varphi) = \frac{2 - k^2}{2\cosh(\alpha) \times \sqrt{1 - k^2}}$$

is satisfied, as shown in App. 2; the factor $\text{sgn}[\vec{A} \times \vec{b}_0^t]$ assumes opposite values on the different roots of last equation [17].

On the contrary when $|\xi| = \frac{\pi}{2}$, that is $\frac{\partial \vec{b}}{\partial J} // \vec{b}_0^t$, the linear contribution of $\chi'$ disappears and the response is purely absorptive [18]. In this case the signal becomes

$$\Delta|\vec{b}_t(B)|^r = r \times \text{sgn}[\vec{A} \times \vec{b}_0^{t\,\wedge}] \times |\vec{b}_0^t|^{r-1} |\vec{A}| \chi'' \quad (22).$$

The scalar amplification can now be defined as $A_r''(J,k) = r \times \text{sgn}[\vec{A} \times \vec{b}_0^{t\,\wedge}] \times |\vec{b}_0^t|^{r-1} |\vec{A}|$ whose explicit expression, taking into account Eq. 16, results

$$A_r''(J,k) = r \times \text{sgn}\left[\vec{b}_0^t \times \frac{\partial \vec{b}}{\partial J}\right] 4\pi^2 \eta \frac{h}{m} \frac{e^{-\alpha}}{1 - e^{-2\alpha}} |\vec{b}_0^t|^{r-1} \left(1 - |\vec{b}_0|^2\right) \sqrt{1 - |\vec{b}_{c1}|^2} \quad (23).$$

In absence of the reference arm the condition $\frac{\partial \vec{b}}{\partial J} // \vec{b}_0^t$ is verified if and only if $\delta\varphi = 0$, as shown in App. 2.

Due to its practical relevance, the purely absorptive response in linear regime deserves particular attention; therefore the behavior of $A_r''$ will be investigated in detail in the next section. However it is important to remark that $A_r^a$ represents just the linear part of the sensitivity, as only the first order terms in the expansion of $\left|\vec{b}_t\right|^r$ have been retained.

The formal equivalence between the expressions of $A_r^a$ and $A_r^\phi$ allows a similar analysis for the purely dispersive linear response.

### VI. Absorptive linear response

As shown above, the absorptive response in linear regime can be discussed in terms of the scalar amplification $A_r^a$. The main difference between $A_r^a$ and the modulus of its complex counterpart $\left|\vec{A}\right|$ is the presence of the term $\left|\vec{b}_0^t\right|^{r-1}$, which in general has relevant effects on the linear response near the point of ideal compensation $\vec{b}_0^t \circ \vec{b}_0 \sqrt{1 - \left|\vec{b}_{c1}\right|^2} + \vec{b}_{c2} = 0$. Moreover, when the point $\vec{b}_0^t = 0$ is crossed the factor $\text{sgn}\left[\vec{b}_0^t \times \frac{\partial \vec{b}}{\partial J}\right]$ leads to a lineshape inversion.

The behavior of $A_r^a$ strictly depends on the responsivity of the employed detector, as shown in the following; the effect of the reference arm will also be outlined.

*Linear Detector*

For a linear detector ($r=1$) the scalar amplification becomes

$$A_1^a(J,k) = \text{sgn}\left[\vec{b}_0^t \times \frac{\partial \vec{b}}{\partial J}\right] 4\pi^2 n \frac{h}{mc} \frac{e^{-\alpha}}{1 - e^{-2\alpha}} \left(1 - \left|\vec{b}_0\right|^2\right) \sqrt{1 - \left|\vec{b}_{c1}\right|^2} =$$

$$= \text{sgn}\left[\vec{b}_0^t \times \frac{\partial \vec{b}}{\partial J}\right] \times \left|A(J,k)\right| \quad (24),$$

which is, as expected, very similar to the complex amplification defined in Eq. 16; as a consequence the reference arm reduces the sensitivity of the spectrometer, at least in what concerns its coherent response.

Assuming $\vec{b}_{c1} = \vec{b}_{c2} = 0$ Eq. 24 becomes, at the first order in $\alpha$,

$$A_1^a(\alpha,k) = \text{sgn}[b_0] \frac{2\pi}{mc} (1 - b_0^2) Q_0 h \quad (25),$$

being in this case $b_0$ real and $\frac{\partial b}{\partial J}$ real and positive.

*Superlinear detector*

More differences between $\left|\vec{A}\right|$ and $A_r^a$ arise when superlinear detectors ($r>1$) are considered. In this case $A_r^a$ goes to zero at the point $\vec{b}_0^t = 0$, where the linear sensitivity reaches a minimum; the maximum of the sensitivity can be obtained by inspection of Eq. 23.

Among the superlinear detectors, the most important is the quadratic one, corresponding for instance to a typical bolometer; in this case Eq. 23 becomes

$$A_2^a(J,k) = 2 \times \text{sgn}\left[\vec{b}_0^t \times \frac{\partial \vec{b}}{\partial J}\right] 4\pi^2 n \frac{h}{mc} \frac{e^{-\alpha}}{1 - e^{-2\alpha}} \left|\vec{b}_0^t\right| \left(1 - \left|\vec{b}_0\right|^2\right) \sqrt{1 - \left|\vec{b}_{c1}\right|^2} \quad (26).$$

When the reference arm is not used $A_2^a$ becomes, at the first order in $\alpha$,

$$A_2^a(\alpha,k) = 2 \times \text{sgn}[b_0] \frac{2\pi}{mc} |b_0| (1 - b_0^2) Q_0 h \quad (27).$$

The spectrometer then shows a vanishing linear response near the critical coupling condition, where the energy density on the sample is maximum. The maximum value of the linear sensitivity can be calculated by imposing $\frac{\partial}{\partial |b_0|}\left[|b_0|\times\left(1-|b_0|^2\right)\right]=0$, which gives $|\bar{b}_0|^2=\frac{1}{3}$ and $A'_{2,\max}=\frac{4}{3\sqrt{3}}\frac{2\pi}{mc^2}Q_0\eta$; this result, analogous to that quoted by Feher [7], is independent on the other parameters of the resonator.

The correspondence between a minimum in the linear sensitivity and the maximum in the energy density is a peculiarity of any spectrometer based on superlinear detectors. The introduction of a reference arm can avoid this effect; indeed, assuming $|\vec{b}_{c1}|=|\vec{b}_{c2}|$ the linear amplification for quadratic detectors in the case of critical coupling becomes

$$A'_2(\alpha,k_c)=2\times\mathrm{sgn}[b_{c2}]\frac{2\pi}{mc^2}|b_{c2}|\sqrt{1-|b_{c2}|^2}\,Q_0\eta,$$

whose maximum is obtained for $|b_{c2}|^2=\frac{1}{2}$; the corresponding value $A'_{2,\max}=\frac{2\pi}{mc^2}Q_0\eta$ equals that found for a linear detector in the case of critical coupling (see Eq. 25).

The last case to be discussed is that of a hypothetical sublinear detector ($r<1$); in this case the amplification $A'_r$ diverges at the ideal compensation point $\vec{b}_0^t=0$, so that the maximum of sensitivity is expected near this region.

## VII. Conclusions

In conclusion, a vectorial analysis of the response of a magnetic resonance spectrometer, based on traveling wave resonators and including the reference arm and the AFC system, has been developed. This analysis allows for exploiting the potentialities of the vectorial detection, which has been already proved effective [6, 19], and appears very general being valid also for stationary wave resonators. Unlike the usual RLC lumped circuit representation, the proposed approach is based on the analysis of the attenuation and the dephasing of the wave in the resonator; in this way the response of the spectrometer is obtained in terms of a complex resonance frequency, which can be measured in any working condition. Moreover this representation can be extended to the analysis of perturbing samples, provided that the resulting complex resonance frequency and fields distribution can be calculated.

The obtained complex response has been specialized to the use of scalar detectors, which can lead to distorted lineshapes in spite of the assumed linear vectorial response; the developed approach will allow for extracting the intrinsic lineshape of the sample in any working condition through a simple graphical analysis. Finally, the knowledge of the response function for arbitrary excitation frequency allows the analysis of more complicated setups, as in the case of bistability phenomena in electron paramagnetic resonance experiments [20].

Practical applications of the proposed model will be discussed and experimentally verified on a spectrometer based on whispering gallery resonators in a forthcoming paper.

# App. 1 – Properties of b and $\frac{\partial b}{\partial J}$

The real and imaginary parts of $\vec{b}$ and $\frac{\partial \vec{b}}{\partial J}$ are given by

$$\mathrm{Re}[b] = \frac{\sqrt{1-k^2}(1+e^{-2a}) - e^{-a}(2-k^2)\cos j}{1+(1-k^2)e^{-2a} - 2\sqrt{1-k^2}e^{-a}\cos j},$$

$$\mathrm{Im}[b] = \frac{e^{-a}k^2 \sin j}{1+(1-k^2)e^{-2a} - 2\sqrt{1-k^2}e^{-a}\cos j},$$

$$\mathrm{Re}\left[\frac{\partial b}{\partial J}\right] = k^2 e^{-a} \frac{\cos j \times [1+(1-k^2)e^{-2a}] - 2\sqrt{1-k^2}e^{-a}}{[1+(1-k^2)e^{-2a} - 2\sqrt{1-k^2}e^{-a}\cos j]^2}$$

$$\mathrm{Im}\left[\frac{\partial b}{\partial J}\right] = k^2 e^{-a} \frac{\sin j \,[(1-k^2)e^{-2a} - 1]}{[1+(1-k^2)e^{-2a} - 2\sqrt{1-k^2}e^{-a}\cos j]^2}.$$

The latter equation implies that $\mathrm{Im}\left[\frac{\partial b}{\partial J}\right] = 0$ when $\sin j = 0$; being $\delta j \ll 1$, it follows that $\frac{\partial b}{\partial J}$ is real for $\delta j = 0$, for any allowed value of $a$ and $k$.

On the other side, $\mathrm{Re}\left[\frac{\partial b}{\partial J}\right] = 0$ corresponds to $\cos(\delta j) = \frac{2\sqrt{1-k^2}\,e^{-a}}{1+(1-k^2)e^{-2a}}$. This latter equation gives real roots for any allowed value of $k$ and $a$.

# App. 2 - Crossed relations between $\vec{b}$ and $\frac{\partial \vec{b}}{\partial J}$

Consider now the conditions $\vec{b} \wedge \frac{\partial \vec{b}}{\partial J}$ and $\vec{b} \parallel \frac{\partial \vec{b}}{\partial J}$.

The vectors $\vec{b}$ and $\frac{\partial \vec{b}}{\partial J}$ are parallel when the corresponding complex numbers have the same phase, that is when the ratio between these numbers is real; otherwise, they are perpendicular when this ratio is an imaginary number. In the present case the complex quantities to be compared are $\frac{\sqrt{1-k^2} - e^{-J}}{1 - \sqrt{1-k^2}\times e^{-J}}$ (Eq. 1) and $\frac{k^2 e^{-J}}{(1-\sqrt{1-k^2}\,e^{-J})^2}$ (Eq. 7).

Their ratio becomes, neglecting the real factor $k^2$,

$$\frac{(\sqrt{1-k^2}-e^{-J})\times(1-\sqrt{1-k^2}\times e^{-J})}{e^{-J}} = \sqrt{1-k^2}(e^J + e^{-J}) + k^2 - 2 =$$

$$= \sqrt{1-k^2}[\cos j\,(e^a + e^{-a}) + i\times\sin j\,(e^a - e^{-a})] + k^2 - 2 \qquad (A1.1);$$

this expression is real when $\sin j\,(e^a - e^{-a}) = 0$, that is for $\sin j = 0 \;\hat{\cup}\; \delta j = 0$ for any allowed value of $a$ and $k$, as shown in App. 1.

On the contrary, Eq. (A1.1) becomes an imaginary quantity when

$$\cos(\delta j) = \frac{2-k^2}{2\cosh(a)\times\sqrt{1-k^2}} \qquad (A2.1).$$

Being $\delta j \ll 1$, this equation has a pair of opposite roots. The second member of Eq. A2.1 is an increasing function of $k$, which assumes the unitary value at the critical coupling, so that real roots can be found $\forall \alpha$ and $\forall k \leq k_c$.

# REFERENCES

1) J. Yu, X.S. Yao, and L. Maleki, High-Q whispering gallery mode dielectric resonator bandpass filter with microstrip line coupling and photonic bandgap mode-suppression, *IEEE Microwave Guided Wave Lett.* **10**, 310-12 (2000).

2) I.M. Tobar, J.G. Hartnett, E.N. Ivanov, D. Cros, P. Blondy, and P. Guillon, Cryogenically cooled sapphire-rutile dielectric resonators for ultrahigh-frequency stable oscillators for terrestrial and space applications (atomic frequency standards), *IEEE Trans. Microwave Theory Tech.* **48,** 1265-1269 (2000).

3) T. Baba, H. Yamada, and A. Sakai, Direct observation of lasing mode in a microdisk laser by a near-field-probing technique, *Appl. Phys. Lett.* **77**, 1584-1586 (2000).

4) G. Annino, M. Cassettari, M. Fittipaldi, I. Longo, M. Martinelli, C.A. Massa, and L.A. Pardi, High-field, multifrequency EPR spectroscopy using whispering gallery dielectric resonators, *J. Magn. Reson.* **143**, 88-94 (2000).

5) G. Annino, M. Cassettari, M. Fittipaldi, M. Martinelli, Analysis of EPR lineshapes for systems interacting with whispering gallery mode resonators, *Book of Abstracts, 30th Congress AMPERE on Magnetic Resonance and Related Phenomena*, Lisbon, Portugal, 23-28 July 2000, P6.

6) M. Mola, S. Hill, P. Goy, and M. Gross, Instrumentation for millimeter-wave magnetoelectrodynamic investigations of low-dimensional conductors and superconductors, *Rev. Sci. Instrum.* **71**, 186-200 (2000).

7) G. Feher, Sensitivity considerations in microwave paramagnetic resonance absorption techniques, *Bell Syst. Tech. J.* **36**, 449-484 (1957).

8) T.H. Wilmshurst, W.A. Gambling, and D.J.E. Ingram, Sensitivity of resonant cavity and travelling-wave E.S.R. spectrometers, *J. Electron. Control* **13**, 339-360 (1962).

9) P. Ehrenfest, Adiabatic invariants and the theory of quanta, *Phil. Mag.* **33**, 500-513 (1917).

10) M. Muraguchi, K. Araki, and Y. Naito, A new type of isolator for millimeter-wave integrated circuits using a nonreciprocal traveling-wave resonator, *IEEE Trans. Microwave Theory Tech.* **30**, 1867-1873 (1982).

11) D.R. Rowland and J.D. Love, Evanescent wave coupling of whispering gallery modes of a dielectric cylinder, *IEE Proc. J.* **140**, 177-188 (1993).

12) G. Annino, M. Cassettari, M. Fittipaldi, and M. Martinelli, Complex response function of whispering gallery dielectric resonators, *Int. J. Infrared Millim. Waves* **22**, 1485-1494 (2001).

13) H.A. Atwater, "Introduction to Microwave Theory", McGraw-Hill Book Company, New York (1962).

14) G. Annino, D. Bertolini, M. Cassettari, M. Fittipaldi, I. Longo, and M. Martinelli, Dielectric properties of materials using whispering gallery dielectric resonators:

Experiments and perspectives of ultra-wideband characterization, *J. Chem. Phys.* **112**, 2308-2314 (2000).

15) The relation $\omega_n^c = n \times \omega_{fsr}$ should be substituted with the more accurate version $\omega_n^c = \bar{n} \times \omega_{fsr}$, in which the variation of the free spectral range with the frequency is included, this term being in general non-negligible for dielectric resonators (see Ref. 14). In all the practical cases however $\frac{\bar{n} - n}{n} << 1$, so that the error due to the above approximation is of the second order.

16) C.H. Papas, Thermodynamic considerations of electromagnetic cavity resonators, *J. Appl. Phys.* **25**, 1552-1553 (1954).

17) $sgn[\vec{A} \times \vec{b}_0]$ is positive when the two vectors $\vec{b}_0$ and $\vec{A}$ are parallel, and negative when they are antiparallel. Each of the two conditions is verified for a root, $\delta\varphi_1$ or $\delta\varphi_2$ of the Eq. A2.1, which satisfy $\delta\varphi_1 = -\delta\varphi_2$; indeed, from the parity of $Re[\vec{b}]$, $Im[\vec{b}]$, $Re\left[\frac{\partial \vec{b}}{\partial \varphi}\right]$ and $Im\left[\frac{\partial \vec{b}}{\partial \varphi}\right]$ in respect to $\delta\varphi$ (see App. 1), it follows that $\vec{A} \times \vec{b}_0$ is an odd function of $\delta\varphi$ and then the dispersive lineshape becomes inverted switching from one root to the other.

18) The previous conditions can be also expressed in terms of partial derivatives with respect to $\alpha$ and $\varphi$, analogously to the analysis developed in Ref. [8]; indeed, the following relations hold: $\frac{\partial |\vec{b}_t|^r}{\partial \alpha} = r |\vec{b}_t|^{r-2} \sqrt{1 - |b_{c1}|^2} \frac{\partial \vec{b}}{\partial \varphi} \times \vec{b}_t$ and $\frac{\partial |\vec{b}_t|^r}{\partial \varphi} = r |\vec{b}_t|^{r-2} \sqrt{1 - |b_{c1}|^2} \frac{\partial \vec{b}}{\partial \varphi} \times \vec{b}_t$, so that the response is purely dispersive when $\frac{\partial}{\partial \alpha} |\vec{b}_t|^2 = 0$ and purely absorptive when $\frac{\partial}{\partial \varphi} |\vec{b}_t|^2 = 0$, indipendently from the responsivity of the detector.

19) S. Hill, P.S. Sandhu, M.E.J. Boonman, J.A.A.J. Perenboom, A. Wittlin, S. Uji, J.S. Brooks, R. Kato, H. Sawa, and S. Aonuma, Magnetoelectrodynamics of a three-dimensional organic conductor: observation of cyclotron resonance in $d_2[1,1;0]$-(DMe-DCNQI)$_2$Cu, *Phys. Rev. B* **154**, 13536-13541 (1996).

20) M. Giordano, M. Martinelli, L. Pardi and S. Santucci, Observations of bistability effects in electron paramagnetic resonance experiments, *Phys. Rev. Lett.* **59**, 327-330 (1987).